# Theoretical and experimental comparative study of nonlinear properties of imidazolium cation based ionic liquids

Vinícius Castro Ferreira[a], Letícia Zanchet[b], Wesley Formentin Monteiro[b], Letícia Guerreiro da Trindade[c], Michèle Oberson de Souza[b], Ricardo Rego Bordalo Correia[a]

[a]OPTMA - Optics, Photonics and Materials Group, Institute of Physics, Universidade Federal do Rio Grande do Sul, Porto Alegre, Brazil

[b]LRC - Laboratorio de Reatividade e Catalise, Institute of Chemistry, Universidade Federal do Rio Grande do Sul, Porto Alegre, Brazil

[c]Department of Chemistry, Universidade Estadual Paulista, Bauru, São Paulo, Brazil

**Abstract**

This work describes the experimental and theoretical study of the nonlinear optical properties of the imidazolium cation based ionic liquids and the corresponding thermo-optical parameters. Experimental results of nonlinear optical properties, such as nonlinear refractive index and thermo-optical properties are determined by Z-scan and EZ-scan techniques with a femtosecond laser source. Theoretical simulations of linear and nonlinear optical properties performed by density functional theory (DFT) are discussed, in terms of polarizability, the first and second-order of hyperpolarizability. A correlation between the theoretical and experimental results is presented, where the variation of the signals of each ionic liquid can be compared with their nonlinear optical properties.

# 1. Introduction

Ionic liquids (ILs) also named organic molten salts are composed of ions, a large organic cation, and an anion having a delocalized charge. The cation's asymmetry and flexibility lead to weak ion interactions resulting in compounds with a low tendency to crystallize [1,2]. Since its rediscovery, the ILs have been applied in several branches of chemistry due to a unique combination of physicochemical properties, highlighting a wide electrochemical window, negligible vapor pressure, high electrical conductivity, high ionic mobility, high thermal stability, low flammability and low toxicity [2]. This set of properties, coupled with the possibility of designing ILs by choosing the ions, enables the use of ILs in pharmaceutical and environmental applications [3,4].

In recent years, ionic liquids have had their main characteristics explored, generating promising science and technologies for academic and industrial applications in areas such as catalysis [5–7], materials science [8,9], physical chemistry [10], electrochemistry [11], polymer science [12,13], biology [14], among others, which gives them a multidisciplinary character [15]. The ILs can be part of new technologies that enable to solve current problems faced by society, emerging as effective alternatives.

Among the vast ionic liquids properties that can be studied, optical characterizations are undoubtedly valuable because of the structural and electronic information provided. In the linear optical regime, the optical properties of materials are independent of the light intensity. In contrast, the light field can modify several optical properties in the nonlinear optical regime. The understandings of nonlinear optical (NLO) phenomena are essential to the development of new materials and technology applications, ranging from conversion of wavelength frequencies and optical switches [16] up to ultra-fast laser sources [17] and health sciences [18].

The optical properties of ILs have been widely studied since early 2000. A substance with NLO activity must be polarizable and/or with asymmetric charge distribution and contain a π-conjugated electron moiety, so ILs were recognized as good candidates to be studied in NLO [19]. The ILs present a strong dependence on the properties and nature of the cation and anion. Thus the anionic and cationic moieties' choice may generate materials with optimized properties [20]. Several nonlinear characteristics were studied in these materials, such as fast molecular dynamic processes [21–23]. Only a few studies report data about ILs nonlinear index of refraction, property covered in this work. The variation of the index of refraction generated by thermal effects was evaluated by self-induced thermal effects for different aromatic and alicyclic ILs [24,25]. Most of these values agree with standard interferometric measurements of ILs, as observed for eleven 1-alkyl-3-methylimidazolium-based ionic liquids, registering negative values of the thermo-optical coefficient, dn/dT, in the order of $10^{-4}$-$10^{-3}$ $K^{-1}$. Nonlinear indexes of refraction in the order of $10^{-16}$ $cm^2$ $W^{-1}$ were measured in femtosecond laser systems for colloidal solution of silver nanoparticles dispersed in $BMI.BF_4$ IL [26] and mixtures of azobenzene containing IL crystalline polymer [19], this last one displaying subpicosecond responses. However, to our knowledge, the literature does not describe thermal and electronic effects related to the nonlinear index of refraction of pure ILs exposed to a femtosecond laser source and discuss experimental data associating them to theoretical studies.

In order to study these effects, this paper aimed to evaluate the nonlinear optical behavior of four specific ILs, combining theoretical and experimental approach to prospect the prediction of its optical properties. The four ILs characterized and discussed here are the methylimidazole hydrogen sulfate ($MImH.HSO_4$), butylimidazole hydrogen sulfate ($BImH.HSO_4$), 1-butyl-3-methylimidazolium hydrogen sulfate ($BMI.HSO_4$) and, 1-butyl-3-methylimidazolium trifluoromethane sulfonate ($BMI.CF_3SO_3$). The length and nature of the cation's alkyl chain, as well as the nature of the anion, were modified to verify their influence on the optical properties. Fast and slow response, related to the electronic and thermal effects respectively, was characterized by Z-scan and EZ-scan techniques to those ILs. Theoretical simulation of the Fourier-transform infrared spectroscopy (FTIR) analysis of the ILs was also performed and corroborated with the experimental results.

## 2. Experimental

*2.1 Materials*

The ionic liquids were synthesized using 1-methyl imidazole 99%, (Sigma-Aldrich), 1-chlorobutane 99% (Sigma-Aldrich), ethyl acetate (Sigma-Aldrich) 99.8%, sulfuric acid 98 % (Merck), trifluoromethanesulfonic acid 99% (Sigma-Aldrich), 1-butylimidazole 98% (Sigma-Aldrich).

*2.1 Synthesis and characterization of the ionic liquids*

The 1-butyl-3-methylimidazolium hydrogen sulfate (BMI.HSO$_4$) and 1-butyl-3-methylimidazolium trifluoromethane sulfonate (BMI.CF$_3$SO$_3$) ionic liquids were synthesized by ion exchange reaction from the 1-butyl-3-methylimidazolium chloride (BMI.Cl) IL according to the procedures reported in the literature [27–29]. H$_2$SO$_4$ was added to a solution of BMI.Cl dissolved in deionized water. The resulting solution was maintained under reflux for 4 h at 100 $^o$C and then dried under vacuum at 90 $^o$C. The final compound is a colorless viscous liquid.

BMI.CF$_3$SO$_3$ was synthesized, adding dropwise the trifluoromethanesulfonic acid (CF$_3$SO$_3$H) to an aqueous solution of BMI.Cl placed in an ice bath. After the complete addition of the acid, the solution was kept under stirring for 24 h. Dichloromethane was then used (5 times 50 mL) to extract the IL from the aqueous phase. The organic phase was then washed with deionized water (10 times 100 mL) to remove any chloride salt and acid. Then the IL was dried under vacuum at 120 $^o$C until the complete removal of the residual water and dichloromethane, resulting in yellowish IL.

The two protic ILs, the methylimidazole hydrogen sulfate (MImH.HSO$_4$) and butylimidazole hydrogen sulfate (BImH.HSO$_4$), were produced by reacting a 0.2 mol of sulfuric acid with 1-methylimidazole and 1-butylimidazole respectively at room temperature placed under stirring in an ice bath for 24 h. The resulting solution is heated below 90 °C for removing the solvent leading to the production of a white viscous liquid (MImH.HSO$_4$, 98% yield) and a brown viscous liquid (BImH.HSO$_4$, 93% yield) respectively.

The synthesized ILs were characterized by Fourier-transform infrared spectroscopy (FTIR) using Attenuated Total Reflectance Fourier Transform Infrared Spectroscopy (ATR-FTIR) with a Bruker Alpha-P in the spectral range 4000–500 cm$^{-1}$, and ultraviolet-visible spectrophotometry performed in a Cary 5000 UV-Vis-NIR equipment from Agilent using a 2.0 mm quartz cuvette.

*2.2 Theoretical calculations*

Density functional theory (DFT) study was performed to understand the molecular behavior and structural conformation of the ionic liquids. DFT calculations were achieved using the Becke's three-parameter exchange functional in combination with the Lee, Yang and Parr correlation functional (B3LYP) [30,31] with a 6-311++G(d,p) basis set as implemented in the GAUSSIAN 16 package. Structures were fully optimized, under no symmetry constraints, and vibrational frequency calculations were performed. Final structures have no imaginary frequencies associated with them. The frontier orbital energies were calculated in a single point run in the same theory level. Molecular electrostatic potential maps (MEPs) of total electronic densities using the partial charges were analyzed with Gabedit software [32]. For each ILs, a DFT study was performed for determining the polarizability ($\alpha$), the first order of the hyperpolarizability ($\beta$), and the second-order of hyperpolarizability ($\gamma$), varying the frequencies relations.

*2.3 Nonlinear optical and thermo-optical properties measurements*

The optical and thermo-optical properties of the ionic liquids were evaluated by Z-scan and EZ-scan techniques for the determination of the thermo-optical coefficient, nonlinear absorption and nonlinear refractive index (also called intensity-dependent refractive index). Z-scan is a widely used technique that characterizes the nonlinear refractive index, nonlinear absorption/saturation, and thermo-optical properties by the variation of wavefront curvature [33]. The eclipse configuration, EZ-scan, can measure the same properties of the Z-scan technique but with better sensitivity and signal-noise ratio [34]. The difference between them is in the spatial filter used to select the part of the beam, which will be detected: Z-scan uses an aperture while EZ-scan uses a disc. In both cases, to a high-repetition laser source, cumulative effects will be generated and must be managed to discern among fast (electronic) and slow (thermal) effects [35,36]. A third experimental setup variation that has no spatial filter was also used. In this case, every change of variation of transmittance will come from the nonlinear absorption/saturation in the sample [33]. These configurations were used to characterize the nonlinear optical properties of the ILs. The experimental setup was the same, just changing the spatial filter (disc or aperture) or no filter according to their intended measurement. The light source was a chopped laser beam of a mode-locked Ti: Sapphire laser oscillator (76 MHz, 150 fs, 780 nm), which is focused in a beam waist radius of 35 μm, where the sample is scanned in the z-direction over a range of few Rayleigh lengths ($z_0$). The detection was performed by a fast Si photodiode detector positioned in front of the spatial filter. A second chopper generates a beam modulation at 9 Hz (duty cycle of 2.4% and beam time exposure of 2.66 ms), providing the relaxing time needed for the thermal management procedure [37]. In each sample position, the time evolution of the transmitted signal was recorded by a digital oscilloscope that allows time-resolved analyses. The normalized transmittance to the fast response is acquired at the beginning of the time exposure window, t = 0.

The EZ-scan configuration was used in the characterization of fast response properties, as a nonlinear refractive index. A measurement without spatial filter, open aperture Z-scan, was performed to measure the nonlinear absorption/saturation of the liquids. The Z-scan configuration was used to evaluate the thermal response since this method allows the Z-scan curve to be fitted by an analytical model based on the thermo-optical properties [36].

## 3. Results and discussion

### 3.1 Ionic Liquids characterizations

Fig. 1a shows the experimental Fourier-transform infrared spectroscopy (FTIR) of BMI.CF$_3$SO$_3$, BMI.HSO$_4$, MImH.HSO$_4$ and BImH.HSO$_4$, and the corresponding theoretical spectra performed for this study (Fig. 2b). The results show that experimental and theoretical data converge. The BMI.CF$_3$SO$_3$ IL spectrum (Fig. 1a) shows the following vibration bands: C-H aromatic at 3173 and 3107 cm$^{-1}$; aliphatic C-H 2968 and 2882 cm$^{-1}$; C-N at 1573 cm$^{-1}$; C=C at 1468 cm$^{-1}$; C-F at 1249 cm$^{-1}$; O=S=O at 1031 cm$^{-1}$ and S-O at 1157 cm$^{-1}$. The band at 853 cm$^{-1}$ is due to the C-H in-plane bending vibration of the imidazolium ring, and the band at 750 cm$^{-1}$ can be assigned to the C-H out-of-plane bending vibration of the imidazolium ring [38,39].

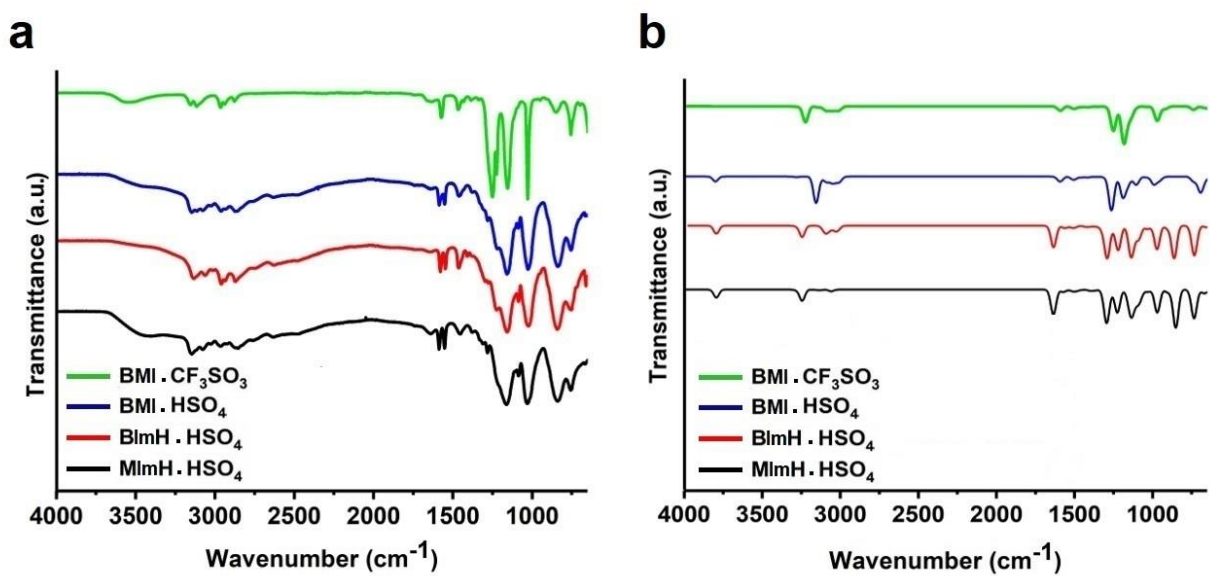

**Fig. 1.** FTIR spectrum of ILs: a) experimental and b) simulated with DFT/B3LYP/6-311++G(d,p).

The spectrum of BMI.HSO$_4$ presents vibrations of imidazolium ring at 3149 and 3105 cm$^{-1}$ corresponding to the C-H bonds, and the bands at 1467 and 1571 cm$^{-1}$ correspond to the C=C and C=N bonds. The band at 837 cm$^{-1}$ is due to C-H in-plane bending vibration of the imidazolium ring, and the band at 753 cm$^{-1}$ can be assigned to the C-H out-of-plane bending vibration of the imidazolium ring. The aliphatic C-H asymmetric and symmetric stretching vibrations of the alkyl chain are represented by the bands at 2962 and 2875 cm$^{-1}$, while the peaks at 1024 and 1229 cm$^{-1}$ are assigned to the S-O and S-OH stretching vibrations of HSO$_4$. The band at 1161 cm$^{-1}$ indicates the stretching vibration of C-N [40,41].

About the MImH.HSO$_4$ and BImH.HSO$_4$ ILs, the bands at 3150, and 3075 cm$^{-1}$ are attributed to the stretching vibrations of the imidazole ring C-H bonds. The aliphatic asymmetric and symmetric (C-H) stretching vibrations of the lateral chain of the imidazole ring appears at 2960 cm$^{-1}$ for MImH.HSO$_4$ and at 2965 and 2930 cm$^{-1}$ for BImH.HSO$_4$. The bands at 1578 and 1454 cm$^{-1}$ in both ionic liquids correspond to C=C and C=N stretching vibrations of the imidazole ring, respectively. The bands at 1222 and 1023 cm$^{-1}$ are assigned to the S-O and S-OH stretching vibrations of HSO$_4$ [40].

The bands at 1162, 840, and 752 cm$^{-1}$ can be attributed to the stretching vibration of C-N, the C-H in-plane bending vibration and the C-H out-of-plane bending vibration of the imidazole ring [42,43]. The simulated spectra (Fig. 1b) of the four ILs structures reported in Fig. 1b, besides being like the experimental spectra do not present imaginary frequencies, *i.e.*, no negative wavenumber. These results confirm that the molecular model obtained theoretically is satisfactory.

Fig. 2 presents the absorbance spectra to the ILs studied. This analysis enables us to choose λ = 780 nm for the laser wavelength value to be used during the study of the ILs optical and thermo-optical properties since, in this region, low UV absorption interference occurs. The determined absorption coefficient values ($a$) related to the laser wavelength used experimentally (λ = 780 nm) for MImH.HSO$_4$ was 0.1± 0.04 cm$^{-1}$, for BImH.HSO$_4$ was 0.05 ± 0.01 cm$^{-1}$, for BMI.HSO$_4$ was 0.02 ± 0.01 cm$^{-1}$ and for BMI.CF$_3$SO$_3$ was 0.04 ± 0.01 cm$^{-1}$.

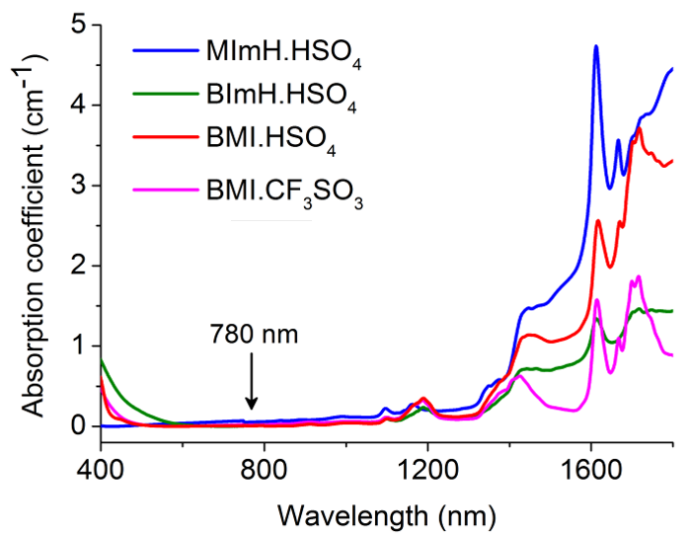

**Fig. 2.** Absorption spectra of the ILs studied. The arrow pointing at λ = 780 nm indicates the absorption coefficient values of the experimental laser wavelength used in the experimental optical study.

3.2 Theoretical calculations

Fig. 3 presents the DFT results corresponding to the molecular electrostatic potential (MEP) maps obtained for the four ionic liquids (MImH.HSO$_4$, ImH.HSO$_4$ BMI.HSO$_4$ and, BMI.CF$_3$SO$_3$). According to the color scale, the red color present on all the anions' surface indicates nucleophilic domains, while the blue color on the surface of the cation represents their electrophilic character [44].

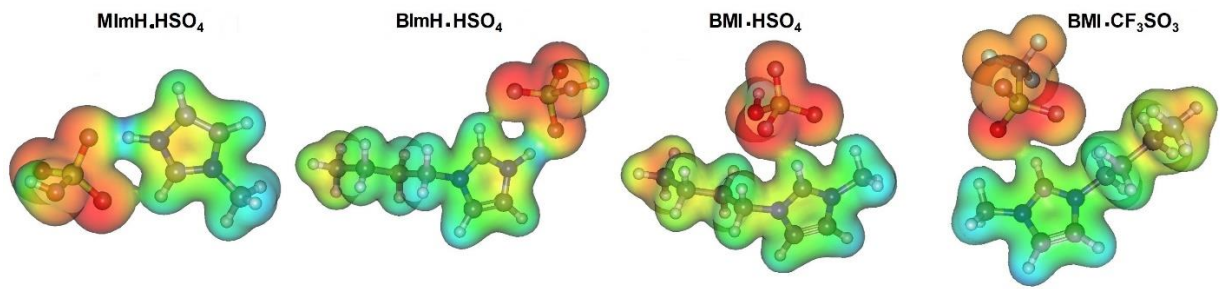

**Fig. 3.** Molecular electrostatic potential maps for ionic liquids. Red color: negative potential; blue color: positive potential.

In the case of the protic ILs (MImH.HSO$_4$ and BImH.HSO$_4$), this study enables to identify the acidic hydrogen of the cation that is interacting highly, probably through an H-bonds with the oxygen atoms of the anion. Both of these ILs present favored anion-cation charge transfer due to this interaction. About the BMI.CF$_3$SO$_3$ and BMI.HSO$_4$ ILs, the HSO$_4$ anion is located at a more centralized place, comparing the structure of MImH.HSO$_4$ and BImH.HSO$_4$. This localization of the anion regarding to the imidazolium ring is probably due to the presence of a methyl group on it.

Fig. 4 shows the results of the frontier molecular orbital (FMO) analysis of the studied ILs. The Highest Occupied Molecular Orbital (HOMO) is nucleophilic and it is considered as an electron donor, while the Lowest Unoccupied Molecular Orbital (LUMO) is usually electrophilic and it is then an acceptor of electron given by a nucleophilic species [45]. The larger orbital volumes represented in Fig. 4 evidence that for all the ILs, the HOMOs are localized in the anion, while the LUMOs are localized in the imidazole ring where the positive charge density is located. The energy difference between the HOMO and LUMO energy (energy gap) is associated with the charge transfer interaction within the frontier orbitals. Comparing the gap energy obtained for all the ILs, the lowest one is observed for the BMI.HSO$_4$ IL. This result can be attributed to the presence of the methyl group in the imidazolium ring that generates an increase in the HOMO and a decrease in the LUMO values. This reduced gap energy allows the transfer of electrons from the anion to the cation.

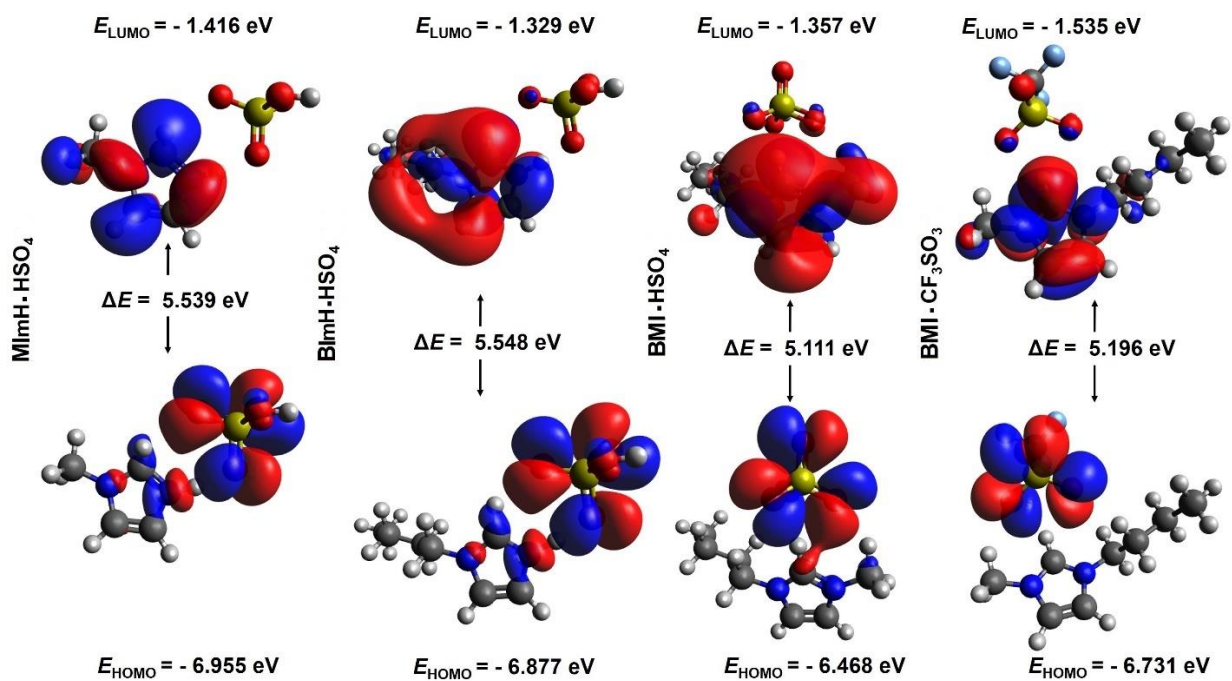

**Fig. 4.** Atomic orbital HOMO - LUMO composition of the frontier molecular orbital of ILs.

Fig. 4 also shows that the HOMO orbitals have the lowest electron distribution in the atoms that have the shortest distances of interaction. This correlation is the most significant in the case of MImH.HSO$_4$ and BImH.HSO$_4$ that are protic ionic liquids. This result can be associated with an important degree of intramolecular charge transfer from the anion to the cation [46]. The interaction energies of ion-pairs are defined as follows:

$$E_{(kJ\,mol^{-1})} = 2625.5\,[E_{is} - (E^+ + E^-)] \qquad (1)$$

where $E_{is}$ is the energy of the ionic system, and $E^+$ and $E^-$ are the energy of the cation and anion, respectively. The interaction energies values obtained for MImH.HSO$_4$, BImH.HSO$_4$, BMI.HSO$_4$ and BMI.CF$_3$SO$_3$ were -412.20 kJ.mol$^{-1}$, -401.70 kJ.mol$^{-1}$, -360.20 and -339.21 kJ.mol$^{-1}$. The comparison of the values obtained for the two protic

ILs (BImH.HSO$_4$ and MImH.HSO$_4$) shows that the increase in the alkyl chain of the imidazole ring leads to an increase in the interaction energy, indicating a favorable charge transfer between the ions [46].

The lowest interaction energy values obtained for BMI.HSO$_4$ and BMI.CF$_3$SO$_3$ indicates that the presence of a methyl group in the imidazolium ring increases the interaction between the cation and the anion. This result can be corroborated with the results obtained through the MEPs study (Fig. 3), which showed that the anion HSO$_4$ was located nearest to the imidazolium ring center in comparison with the protic ILs. A correlation between the interaction energies and the HOMO-LUMO energy gap can then be made.

The dipole moment ($\mu$) in a molecule infers the molecular charge distribution, and its determination enables the study of intermolecular interactions [46]. The polarizability ($\alpha$) and the first hyperpolarizability ($\beta$) of a compound, that are associated with the intramolecular charge transfer, result from the electron cloud movement, for example, from an electron donor group to an acceptor group through the π - conjugated framework. Results obtained with DFT are directly related to the linear and nonlinear optical properties. The polarizability is proportional to the static or the frequency-dependence polarizability, while $\Delta\alpha$ represents the anisotropy of polarizability, i.e., the variation between the polarizability value obtained on the axis of the polarizability and a perpendicular axis to the polarizability axis. The first order of hyperpolarizability, ($\beta$), is proportional to the second-order of susceptibility effects, as second a harmonic generation, optical rectification, electro-optical Pockels effect and three-wave mixing. The second order of hyperpolarizability ($\gamma$) is proportional to the effects of the third order of susceptibility, as four-wave mixing and the responses measured by the Z-scan and EZ-scan techniques.

We evaluated the linear and nonlinear optical properties of the ILs to several optical responses using the wavelength $\lambda$ = 780 nm. Table 1 presents the theoretical values of $\mu$, $\alpha$ and $\beta$, as well the anisotropy of polarizability $\Delta\alpha$ calculated for the studded ILs. The comparison of the $\mu$ values obtained for the protic ILs (MImH.HSO$_4$ and BImH.HSO$_4$) reported in Table 1 shows that the increase in the alkyl chain of the IL (BImH.HSO$_4$) causes an increase in the dipole moment. The presence of the methyl group in the imidazole ring rather than a proton decreases the dipole moment.

**Table 1**

Theoretical values of dipole moment, $\mu$, polarizability, $\alpha$, the anisotropy of polarizability, $\Delta\alpha$, and first hyperpolarizability, $\beta$, for the ionic liquids.

| Sample | $\mu$ (Debye) | $\alpha$ (x 10$^{-24}$ esu) | $\Delta\alpha$ (x 10$^{-24}$ esu) | $\beta$ (x 10$^{-31}$ esu) |
|---|---|---|---|---|
| MImH.HSO$_4$ | 12.47 | 14.08 | 6.66 | 18.37 |
| BImH.HSO$_4$ | 12.96 | 19.64 | 7.96 | 16.42 |
| BMI.HSO$_4$ | 12.55 | 21.29 | 6.95 | 28.51 |
| BMI.CF$_3$SO$_3$ | 13.78 | 22.92 | 7.00 | 31.87 |

The highest first hyperpolarizability values ($\alpha$) are observed for the BMI.HSO$_4$ and BMI.CF$_3$SO$_3$. As this property is dependent on the dipole moment, this result can be associated with the presence of the methyl group in the imidazole ring that promotes a change in the symmetry of the molecule, increasing then the $\alpha$ values [47].

Table 2 presents the polarizability, $\alpha$, and the first order of hyperpolarizability, $\beta$, for different frequencies relations: $\alpha(-\omega; \omega)$ is the frequency-dependent polarizability, $\beta(-\omega; \omega; 0)$ and $\beta(-2\omega; \omega; \omega)$ are related to the electro-optic Pockels effect and the second harmonic generation respectively, both effects from the susceptibility of second-order. Related to the terms of the second-order of hyperpolarizability (Table 2), $\gamma(-\omega; \omega; 0;0)$ presents electro-optic Kerr effect, $\gamma(-2\omega; \omega; \omega; 0)$ presents a DC-induced second harmonic generation or electric field-induced second harmonic. Another response of the second-order of hyperpolarizability, $\gamma(-\omega; \omega; \omega; -\omega)$, is related to the intensity-dependent refractive index or degenerate four-wave mixing, but the DFT software has no library to evaluate this. These three presented terms are all related to the third order of susceptibility. The theoretical results presented are the response of parallel polarization in relation to the applied field.

**Table 2**

DFT results for the polarizability α, and the first order of hyperpolarizability, β, and for the second-order of hyperpolarizability, γ, for different frequencies relations.

| Properties | MImH.HSO$_4$ | BImH.HSO$_4$ | BMI.HSO$_4$ | BMI.CF$_3$SO$_3$ |
|---|---|---|---|---|
| $\alpha(-\omega; \omega)$ ($10^{-24}$ esu) | 14.31 | 19.95 | 21.63 | 23.28 |
| $\beta(-\omega; \omega; 0)$ ($10^{-31}$ esu) | 19.98 | 17.76 | 31.43 | 35.10 |
| $\beta(-2\omega; \omega; \omega)$ ($10^{-31}$ esu) | 24.40 | 21.48 | 40.05 | 44.11 |
| $\gamma(-\omega; \omega; 0;0)$ ($10^{-36}$ esu) | 10.84 | 16.17 | 19.44 | 20.22 |
| $\gamma(-2\omega; \omega; \omega; 0)$ ($10^{-36}$ esu) | 13.94 | 20.30 | 25.76 | 26.30 |

The values of polarizability, the first-order of hyperpolarizability and the second-order of hyperpolarizability are higher for BMI.CF$_3$SO$_3$, followed by BMI.HSO$_4$, indicating that a large chain in the cation, following by the presence of the methyl group in the imidazolium ring presented a stronger influence in these properties. The IL MImH.HSO$_4$, which has the smallest chain in the cation, showed the lowest values of the first-order of hyperpolarizability and the second-order of hyperpolarizability. The unique difference observed is for $\beta(-\omega; \omega; 0)$ ($10^{-31}$ esu) to the IL BImH.HSO$_4$, that is the lowest value comparing all ILs. The imidazolium ring of this IL presented an acid hydrogen that favored charge transfer (see results of Fig. 3 and 4) associated to the cation chain can be related to these values, however, more detailed studies need to be carried out to better explain this phenomenon.

*3.3. Local and nonlocal nonlinear optical characterization*

The fast responses of molecules measured in the refractive index change techniques are usually generated by effects related to the high orders of susceptibility. The fact these effects are also restricted to the spatial concurring presence of the light fields, corresponding to a local response. Examples of these effects are the nonlinear refractive index and nonlinear absorption. Otherwise, thermal effects, which are slower and promote heat diffusion, are not spatially restricted to the beam region, the reason why it corresponds to a nonlocal effect.

The measurements to analyze the fast responses were performed in the EZ-scan configuration. The nonlinear refractive index was evaluated using a model based on literature [34]:

$$n_2 = \frac{\Delta T_{p-v}}{0.68\,(1-S)^{-0,44}\,L_{eff}\,k\,I_0}, \qquad (2)$$

where $\Delta T_{p-v}$ is the peak-valley amplitude of the normalized transmittance variation in the EZ-scan technique, S is the disc blocked factor of the total beam power, $k$ is the wave vector, $I_0$ is the peak intensity and $L_{eff} = (1 - \exp(-aL))/a$ is the sample effective length.

A previously presented procedure of attenuation of systematic errors was employed in these measurements by our group [48]. The scans corresponding to the fast responses obtained with the ILs are presented in the left graphs of Fig. 5. The right side of Fig. 5 presents the results of measurement in the EZ-scan setup 2.3 ms after the chopper aperture corresponding to exposure time to the beam. In these conditions, the thermal effects are predominant. In order to optimize the fast response signal at the beginning of the temporal exposure window, the laser beam's power used was reduced accordingly for each sample. Open aperture Z-scan was also performed to analyze nonlinear absorption/saturation, however, no sample evidenced that sort of response. These results are present inset of Fig. 5. All open aperture measurements were performed with beam peak intensity of 1.7 GW/cm$^2$. Nonlinear refractive indexes of those ILs were determined according to Equation 2 and they are summarized in Table 4. The system was calibrated using a sample of carbon disulfide as the reference. To this sample, the nonlinear refractive index of carbon disulfide was determined as $n_2$ = 2.6 ± 0.3 x 10$^{-15}$ cm$^2$ W$^{-1}$, which correspond with the literature [35,37].

The real part of the third order of susceptibility, $\chi^{(3)}$, is directly related to the degenerate nonlinear refractive index. This term can be evaluated by the equation [49].

$$\chi^{(3)}(-\omega;\omega,\omega,-\omega) = \frac{n_0^2}{283} n_2(\omega) \qquad (3)$$

where $n_0$ is the linear refractive index at the same frequency. Considering that the main contribution would correspond to the electronic contribution of the second-order of hyperpolarizability, $\gamma(-\omega;\omega,\omega,-\omega)$, can be estimated from the correspondent susceptibility and, therefore, by the nonlinear refractive index through the relation considering the local field factor [50]:

$$\gamma = \frac{\chi^{(3)}}{\left[\frac{1}{3}(n_0^2+2)\right]^4 N} \qquad (4)$$

where N is the molecular number density. In this analysis, the linear refractive index used for the estimative ranges from $n_0$ = 1.42 to 1.45 for all liquids [51,52], and this deviation is smaller than the error associated of the measured nonlinear refractive index.

Table 4 presents those estimated hyperpolarizabilities values for each IL. Comparatively, with theoretical results obtained with the DFT theory, these are an order of magnitude smaller. This discrepancy can be associated with a refractive different response, e.g., with the one corresponding to the signal modified by the ILs local structure and the interaction with their vicinities. Based on the value evaluated for the polarization anisotropy, $\Delta\alpha$, one should also consider the optical Kerr effect, which arises from the molecular alignment produced by the optical torque on the induced dipoles [53]. This effect could be evidenced by a technique that discriminated against these effects, as the polarization-resolved Z-scan technique [54]. However, the values of the second-order hyperpolarizability of the first three ILs presented in the table present the same growing tendency verified in the theoretical model.

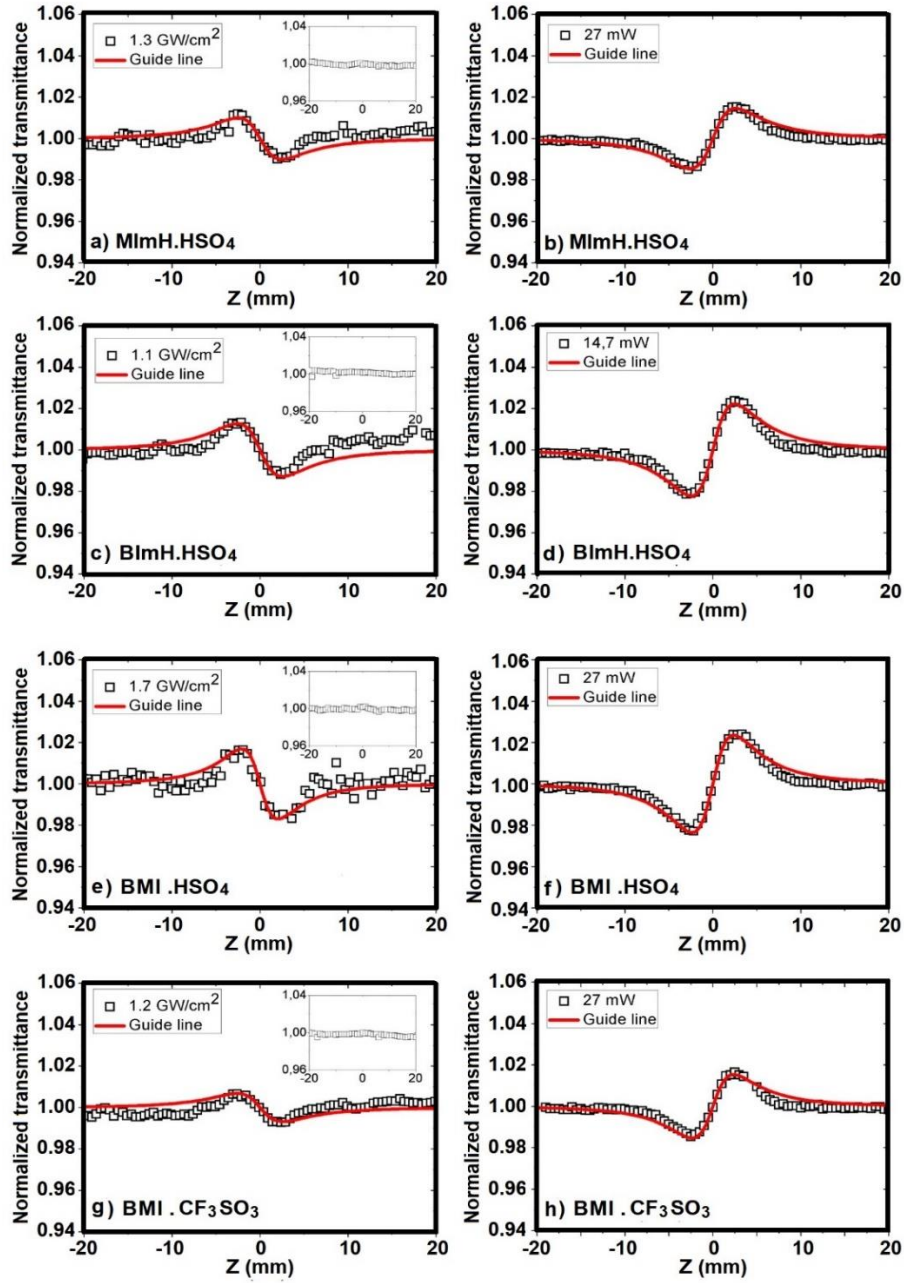

**Fig. 5.** Variation of transmittance registered for the four ILs exposed to a laser beam ($\lambda$ = 780 nm). The left side graphs were registered in the time equals zero (condition for minimizing the thermal effects [36]), while the right-side graphs were recorded after 2.3 ms of beam exposure to identify predominantly the thermal effects. The respectively IL is presented in the Fig. and as well as the intensity and power used in the measurements. Open aperture Z-scan are inset, all them performed with beam peak intensity equals to 1.7 GW cm$^{-2}$.

**Table 4**

Experimental nonlinear optical properties values of the ILs obtained: values of the nonlinear refractive index ($n_2$), third order of susceptibility, $\chi^{(3)}$, and the second order of susceptibility, $\gamma(-\omega; \omega; \omega; -\omega)$.

| Sample | $n_2$ ($10^{-16}$ cm$^2$/W) | $\chi^{(3)}$ ($10^{-22}$ m$^2$/V$^2$) | $\chi^{(3)}$ ($10^{-14}$ esu) | $\gamma(-\omega; \omega; \omega; -\omega)$, ($10^{-36}$ esu) |
|---|---|---|---|---|
| MImH.HSO$_4$ | 2.7 ± 0.7 | 2.0 ± 0.6 | 1.4 ± 0.5 | 0.9 ± 0.2 |
| BImH.HSO$_4$ | 4.2 ± 0.6 | 3.1 ± 0.5 | 2.2 ± 0.3 | 1.6 ± 0.3 |
| BMI.HSO$_4$ | 4.5 ± 0.5 | 3.3 ± 0.4 | 2.4 ± 0.4 | 1.9 ± 0.2 |
| BMI.CF$_3$SO$_3$ | 2.7 ± 0.4 | 2.3 ± 0.4 | 1.6 ± 0.3 | 1.3 ± 0.2 |

Even if a short femtosecond pulse with nanojoules of energy enough not to heat the sample during a single propagation, high repetition rates of these pulses generate cumulative effects, which change the wavefront curvature of the propagating beam. The time evolution of transmittance variation in a high-repetition system was modeled by Falconieri [55]. In the case of linear absorption, that is our situation, the variation of transmittance can be written as:

$$T = 1 + \theta \tan^{-1}\left(\frac{2\tilde{z}}{(9+\tilde{z}^2)t_c/2t + 3 + \tilde{z}^2}\right) \quad (5)$$

where $t_c$ is the thermal characteristic time and normalized scanning parameter $\tilde{z} = z/z_0$. The thermal lens strength, $\theta$, represents the intensity of the lens induced by the sample heating, which curves the beam wavefront and generates the variation of transmittance in this regime. The thermo-optical coefficient, the variation of refractive index (*n*) with the temperature, can be determined by:

$$\frac{dn}{dT} = -\frac{\lambda \kappa}{P \, a \, L_{eff}} \theta \quad (6)$$

where *P* is the laser power in the sample and $\kappa$ is the thermal conductivity, which to ILs can be evaluated applying the model based on equation proposed by Fröba et al. [56]:

$$\kappa = \frac{AM + B}{M\rho} \tag{7}$$

where A = 0:113 g cm$^{-3}$ W m$^{-1}$ K$^{-1}$, and B = 22:65 g$^2$ cm$^{-3}$ W m$^{-1}$ K$^{-1}$ mol$^{-1}$ are constants defined in the temperature of 293.15 K, $\rho$ is the density, and M is the molar mass of the liquid. The thermal conductivity values and uncertainties calculated from the respective densities of each IL studied here are displayed in Table 5.

These properties were estimated to perform a Z-scan analysis with the ILs. The absorption coefficient ($a$), the thermal characteristic time ($t_c$), the thermal lens strength to the laser power equals to 300 mW ($\theta$), the calculated thermal conductivity ($\kappa$) from equation (1) and the thermo-optical coefficient (*dn/dT*), are reported in Table 5 for the four ILs.

**Table 5**

Experimental optical properties values of the ILs obtained: values of the absorption ($\alpha$), thermal characteristic time ($t_c$), thermal lens strength to the laser power equals to 300 mW ($\theta$), thermal conductivity ($\kappa$) and thermo-optical coefficient (dn/dT). The laser beam wavelength was $\lambda$ = 780 nm.

| Sample | $a$ (cm$^{-1}$) | $t_c$ (ms) | $\theta$ | $\kappa$ (W m$^{-1}$K$^{-1}$) | dn/dT (10$^{-6}$ K$^{-1}$) |
|---|---|---|---|---|---|
| MImH.HSO$_4$ | 0.1 ± 0.04 | 1.5 ± 0.2 | 0.3 ± 0.1 | 0.17 ± 0.02 | 5.0 ± 0.8 |
| BImH.HSO$_4$ | 0.05 ± 0.01 | 2.0 ± 0.5 | 1.0 ± 0.3 | 0.15 ± 0.02 | 4.3 ± 0.8 |
| BMI.HSO$_4$ | 0.02 ± 0.01 | 1.7 ± 0.4 | 0.5 ± 0.1 | 0.14 ± 0.02 | 4.6 ± 0.8 |
| BMI.CF$_3$SO$_3$ | 0.04 ± 0.01 | 2.7 ± 0.3 | 0.35 ± 0.1 | 0.13 ± 0.02 | 1.5 ± 0.8 |

All these data were determined for the first time for the MImH.HSO$_4$, BImH.HSO$_4$, BMI.HSO$_4$, and BMI.CF$_3$SO$_3$ ILs. As previously discussed, the literature of thermo-optical properties and nonlinear refractive index involved ILs with a femtosecond laser source are not extensive. Besides that, further studies that relate the electronic effects on experimental results with theoretical simulations were not found.

## 4. Conclusion

Nonlinear optical properties of four ILs, MImH.HSO$_4$, BImH.HSO$_4$, BMI.HSO$_4$, and BMI.CF$_3$SO$_3$, were determined theoretically and experimentally. These four ILs presented a nonlinear refractive index measured in our experimental conditions. Also, the thermo-optical coefficients were determined to all ILs. The DFT analysis could estimate the nonlinear response of several terms of polarizability, first and second order of hyperpolarizability.

The growth of nonlinear values in the DFT analyses of the second order of hyperpolarizability, values are coherent with the growing of nonlinear refractive index measured by the EZ-scan, which can help in a qualitative prediction of ILs before synthesizing and characterize them. One of the ILs did not respect these growing. We believe that, as IL has the biggest difference when compared with the others, the simulations could not reproduce the comparative response. Besides that, no exact agreement was expected, since DFT analysis simulates only a single molecule, not looking at its all vicinity and molecular dynamics responsible for the optical Kerr effect. The technique of polarization-resolved Z-scan measurements will now be introduced to discriminate between electronic and orientational nonlinear effects in these systems. Thermo-optical coefficients were determined to the ILs. Other thermodynamic properties, as thermal characteristic time, thermal lens strength and thermal conductivity, were also evaluated and presented.

**Acknowledgments**

Vinícius Castro Ferreira and Letícia Zanchet thank the finance in part by the CNPq and Letícia Guerreiro da Trindade thanks the finance in part by the Coordenação de Aperfeiçoamento de Pessoal de Nível Superior – Brasil (CAPES). Wesley Formentin Monteiro thanks the CNPq (PDJ process number 157931/2018-8). The authors would like the Theoretical Chemistry Group (TCG/UFRGS) for providing computational resources and Federal University of Rio Grande do Sul (UFRGS).